\documentclass[aps,twocolumn,groupedaddress,showpacs,floats,amssymb]{revtex4}
\usepackage[dvips]{graphicx}
\usepackage{bm}
\begin{document}

\title{Superfluid--Insulator Transition in Commensurate 
Disordered Bosonic Systems:
 Large-Scale Worm-Algorithm Simulations}

\author{Nikolay Prokof'ev}
\author{Boris Svistunov}

\affiliation{Department of Physics, University of
             Massachusetts, Amherst, MA 01003, USA}
\affiliation{Russian Research Center ``Kurchatov Institute",
             123182 Moscow, Russia}



\begin{abstract}

We report results of large-scale Monte Carlo simulations
of superfluid--insulator transitions in commensurate 2D bosonic
systems. In the case of off-diagonal disorder, we find that the
transition is to a gapless incompressible insulator, and its
dynamical critical exponent is $z=1.65 \pm 0.2$. In the case
of diagonal disorder, we prove the conjecture that rare statistical
fluctuations are inseparable from critical fluctuations on the largest
scales and ultimately result in the crossover to the generic universality
class (apparently with $z=2$). However, even at strong disorder,
the universal behavior sets in only at very large space-time distances.
This explains why previous studies of smaller clusters mimicked a
direct superfluid--Mott-insulator transition.

\end{abstract}


\maketitle


Quantum phase transitions in disordered systems remain a poorly
understood phenomenon despite enormous interest in this field. The
$T=0$ transition between the superfluid (SF) and insulating (I)
phases is believed to determine properties of various condensed
matter systems: $^4$He in porous media and aerogels
\cite{Crowell9597,Reppy2000}, $^4$He on various substrates
\cite{quartz1,Crowell9597,hydrogen1}, thin superconducting films
\cite{Liu,Valles,Hebard,Yazdani,Goldman,Kravchenko,Sarachik},
Josephson-junction arrays \cite{Zant}, disordered magnets
\cite{Read,Vajk}, etc.

 There are strong arguments that the basic Hamiltonian which
captures all the essential physics of the SF-I transition  is the
bosonic Hubbard model with disordered chemical potential
\cite{Ma8586,Fisher89,Fisher91,Wallin94}. In the limit of large
occupation numbers, the bosonic Hubbard Hamiltonian is equivalent
to the system of coupled Josephson junctions.
Fermionic systems map to this Hamiltonian under the assumption
that Cooper pairs  are preformed at finite temperature, and the 
transition is driven only by quantum fluctuations of the phase 
of the complex order parameter. To deal with granular superconductors
one may also introduce disorder to hopping amplitudes.

It was suggested in Ref.~\cite{Fisher89} that one has to consider
only two competing insulating phases---the incompressible (gaped)
Mott-insulator phase (MI), and  the  compressible gapless Bose
glass (BG) phase. However, more recently it was argued that apart
from the BG phase characterized as a compressible insulator with
variable-range-hopping conductivity at finite temperature
\cite{Fisher89}, there may exist other phases such as a Bose metal
with finite conductivity  in the $T \to 0$ limit \cite{Phillips}
and an incompressible  Mott glass with the conductivity pseudogap
\cite{Giarmarchi}. Theoretical calculations for the strongly
coupled SF-I  critical point are notoriously difficult and are not
based on  well controlled approximations since localization and
interaction effects cannot be separated \cite{Mukh,Igor01}. Thus
even the qualitative understanding of the phase diagram  is still
under debate. In particular, it was argued in
\cite{Fisher89,Fisher91,Wallin94,Wen2,FM96} that MI and SF phases
are always separated by the BG phase at any  finite disorder.
However experiments \cite{Crowell9597},  most Monte Carlo
simulations \cite{Krauth91,Pai,Kisker}, and  other theories
\cite{Zhang,Singh,Pazmandi} present  evidence in favor of a direct
transition between MI and SF  phases (in the case of commensurate
filling of the lattice and not so strong disorder). In $d=1$ this
contradiction was apparently resolved using arguments based on
rare statistical fluctuations \cite{FM96}
(Lifshitz-Griffiths-McCoy singularities
\cite{Lifshitz,Griffiths,Coy}), renormalization-group equations
\cite{Svistunov96,Igor98a}, and quantum Monte Carlo simulations
\cite{ourGlass}).

In this Letter, we numerically address the problem of the
SF-I transition in a disordered commensurate 2D system. Our
large-scale simulations based on the classical Worm Algorithm
\cite{Worm} demonstrate
the absence of the direct SF-MI transition. We clearly see, however,
that---even at strong disorder---the universal asymptotic
long-range behavior sets in only at large space-time distances.
This result, on one hand, explains previous observations of the direct
SF-MI transitions in simulations of much smaller
clusters, and, on the other hand, implies that the superfluid
stiffness and compressibility should obey generic scaling laws
only in a very close vicinity of the phase transition point
which may be hard to study experimentally.

The Worm algorithm (WA) \cite{Worm} is a high-performance
universal Monte Carlo scheme applicable to any model with the
configuration space of continuous paths. The principle of WA is to
work in an {\it extended} configuration space combining the physical
closed-path sector and the broken-path sector with two path endpoints.
{\it All} updates in the extended configuration space are through the motion
of the endpoints (or even just one of them), so that the configuration
evolution looks like a creeping worm.
Though the WA updating strategy is based on local Metropolis moves
(which is a key issue for its universality),
it has a remarkable efficiency equivalent to that of the best
cluster algorithms  \cite{worm2001}.

WA was introduced initially for the continuous-time path integral,
or worldline, representation of lattice quantum systems
\cite{Worm}. More recently it was implemented within the
Stochastic Series Expansion (SSE) method \cite{SSE}, and
generalized to classical lattice systems in the closed-path
representations \cite{worm2001}. The optimal choice of WA depends
on the problem being addressed; e.g. the quantum worldline scheme
is best suited for {\it ab initio} simulations of bosons in
optical lattices \cite{ourOlattice}, while the SSE scheme has
certain storage and computational advantages when simulating
quantum models with a restricted Hilbert space, like spins and
hard-core bosons \cite{Sandviklast,DT}. However, if one is
interested in generic properties of quantum phase transitions,
then the optimal choice is a $(d+1)$-dimensional {\it classical}
scheme, which is algorithmically superior from all points of view.
This approach was advocated in Ref.~\cite{Wallin94}, and most
recently, using WA, in Ref.~\cite{AS}.

The easiest way to derive the classical $(d+1)$-dimensional
equivalent of the $d$-dimensional quantum model is to start from
the lattice path integral for particle trajectories in imaginary
time. The basic step is to ``roughen'' trajectories by replacing
integrals over kink positions (particle jumps between lattice
sites) in time by discrete sums. To this end one introduces a grid
of imaginary times and requires that kinks occur only at time
slices forming the grid. Bonds in the spatial direction of thus
obtained $(d+1)$-dimensional lattice carry an integer charge---a
spatial current equal to an algebraic sum of kinks associated with
a given pair of sites and a time slice (the kink sign specifies
its direction). The finite value of the time interval between
slices, $\Delta \tau$, makes the roughening procedure, generally
speaking, not unique. Only in the physical limit of $\Delta \tau
\to 0$ the absolute values of spatial currents are either zero or
unity. At strong roughening the constraint that the absolute value
of the spatial bond charge is $\le 1$ is no longer necessary. The
maximal value now depends on how we roughen the original
trajectories with more than one kink between a given pair of sites
during the time interval $\Delta \tau$---either as irrelevant rare
events that are neglected, or by ascribing all such kinks to the
same spatial bond of the $(d+1)$-dimensional lattice. Note, there
is no {\it qualitative} difference between the two cases, and 
the choice is just a matter of convenience. Bonds in the temporal 
direction also carry an integer
charge, equal to the occupation number of a given site between
adjacent time slices. For the sake of symmetry, it is convenient
to refer to the temporal-bond charges as temporal currents, so
that each bond of the $(d+1)$-lattice carries an integer current,
$J$. The conservation of the particle number imposes an obvious
constraint on bond currents---the divergence of the
$(d+1)$-current at any $(d+1)$-site is zero. Graphically, if one
represents the bond currents by oriented lines (in accordance with
the sign of $J$) then all contributions to the partition-function
will have a form of closed-path configurations of such lines,
while configurations for the Green-function will have two
endpoints. WA simulates both quantities simultaneously, by
switching between partition-function and Green-function sectors
\cite{Worm,worm2001}.

Let us denote by $\{ J_{{\bf x}, \alpha } \}$ the bond current
configuration where ${\bf x} = ({\bf r}, \tau )$ are discrete
space-time coordinates, and index $\alpha = \hat{r}_1, \ldots ,
\hat{r}_d, \hat{\tau}$ stands for unit vectors of axis directions,
so that $({\bf x}, \alpha )$ defines a bond in the direction
$\alpha$, adjacent to the site ${\bf x}$. As usual, the
configuration weight $W[\{ J_{{\bf x}, \alpha}\} ]$ may be
formally written as a Gibbs factor $\exp \{ -H/T \}$ which for
positive-definite $W$ defines the classical bond Hamiltonian $H/T
= - \ln W$. For models with on-site particle-particle interactions
the configuration weight is simply given by the product of bond
weights $W_{{\bf r} \alpha} ( J_{{\bf x}, \alpha} )$, and,
correspondingly,
\begin{equation}
H = \sum_{{\bf x} \alpha }  \; H_{{\bf r} \alpha} (J_{{\bf x},
\alpha}) \;. \label{H}
\end{equation}
The zero-divergence constraint can be written as $\sum_{\alpha}
J_{{\bf x}, \alpha} + \sum_{\alpha}  J_{{\bf x}, -\alpha} = 0$,
where, by definition, the direction $-\alpha$ is understood as
opposite to $\alpha$ and $J_{{\bf x}, -\alpha}=-J_{{\bf x}-\alpha,
\alpha}$.

\begin{figure}[tbp]
\includegraphics[width=6.5cm]{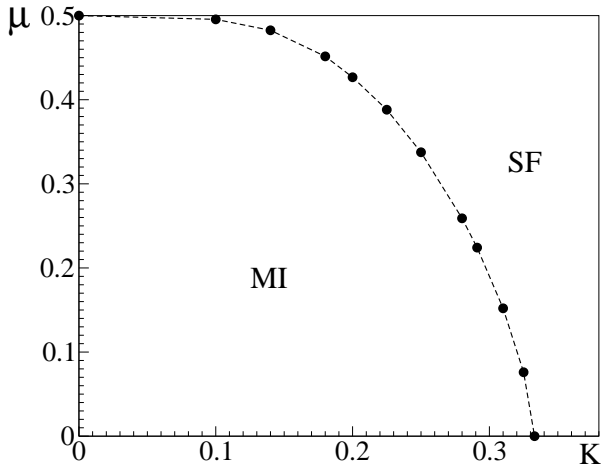}
\vspace*{-2.cm} \caption{ The phase diagram of model (\ref{model})
in the absence of disorder. Error bars are of order $10^{-3}$ and
smaller than point size. } \label{Fig0}
\end{figure}

In this paper, we are interested only in the universal critical
behavior of the model (\ref{H}), which is supposed to be
insensitive to quantitative details of the Hamiltonian. This
freedom may be used to make spatial and temporal
directions symmetric with respect to each other. In the original quantum
model, the temporal direction is not symmetric even with respect
to its opposite because occupation numbers are always positive.
Nevertheless, one may count them from some integer $n_0$,
substitute $J_{{\bf x}, \hat{\tau}} \to J_{{\bf x}, \hat{\tau}} -
n_0$, and formally consider  $J_{{\bf x}, \hat{\tau}} \in (-\infty
, \infty)$. Alternatively, one may introduce a symmetric
constraint, say, $|J| \leq 1$. The latter case, reminiscent of the
mapping between bosonic and spin-1 systems, seems to be more
natural and computationally economic. Historically, however, a
model with $J \in (-\infty , \infty)$, motivated by its derivation
from the Josephson-junction array Hamiltonian \cite{Wallin94}, was
adopted. To retain the possibility of a direct comparison with
previous numeric studies, we work with the same  model
\cite{Wallin94}:
\begin{equation}
H/T =  \sum_{{\bf x} \alpha } \left[ {1 \over 2}
J^2_{{\bf x}, \alpha} - \delta_{\alpha, \hat{\tau}} \mu_{\bf
r} J_{{\bf x} , \hat{\tau}} \right] / K \;.
\label{model}
\end{equation}
In terms of the underlying bosonic system, $K$ represents the
particle hopping amplitude in units of the on-site repulsion, the
discrete field $\mu_{\bf r}= \mu_0 + \tilde{\mu}_{\bf r}$ involves
the chemical potential $\mu_0$ and the white-noise diagonal
disorder $\tilde{\mu}_{\bf r} \in [-\Delta,\Delta]$. In this study
we are concerned with the commensurate filling of the lattice,
i.e. $n = \langle \! \langle J_{{\bf x} , \hat{\tau}} \rangle \!
\rangle =$integer, where $\langle \! \langle \dots \rangle \!
\rangle$ stands for the average over all lattice points,
statistical and disorder fluctuations, and thus set $\mu_0 =0$.
[An accurate study of the half-integer $n$ case has been reported
recently by Alet and S{\o}rensen \cite{AS}]. We also consider
model (\ref{model}) with the off-diagonal disorder introduced by
letting the parameter $K$ to be dependent on ${\bf r}$ and spatial
direction, and confine ourselves to the case of a broken-bond
disorder, where for some randomly chosen ${\bf r}$ and $\alpha' =
\hat{r}_1, \dots\hat{r}_d$ we set $K_{{\bf r} \tau \alpha'} \to 0$
(equivalent to a rigid constraint $J = 0$ on the corresponding
bond). The phase diagram of the homogeneous system is shown in
Fig.~\ref{Fig0}; the SF-I transition at the commensurate filling
is located at $K_c^{(0)}=0.33305(5)$ \cite{AS}. The value of the
MI gap is defined as half the difference between the critical
values of the chemical potential in Fig.~\ref{Fig0}, i.e. $2E_{\rm
gap} (K) = \mu_c^{\rm (up)} - \mu_c^{\rm (down)}$.

\begin{figure}[tbp]
\includegraphics[width=6.5cm]{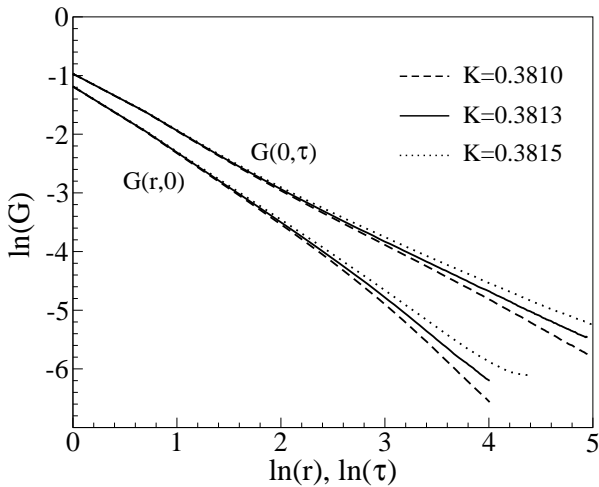}
\vspace*{-2.cm} 
\caption{ Correlation functions $G(r,0)$ and $G (0,\tau )$ for 
$K=0.3810,~0.3813,~0.3815$ and system size $L \times L \times L_{\tau} =
160 \times 160 \times 500 $. Error bars are comparable to the line 
widths.
}
\label{Fig1}
\end{figure}
\begin{figure}[tbp]
\includegraphics[width=6.5cm]{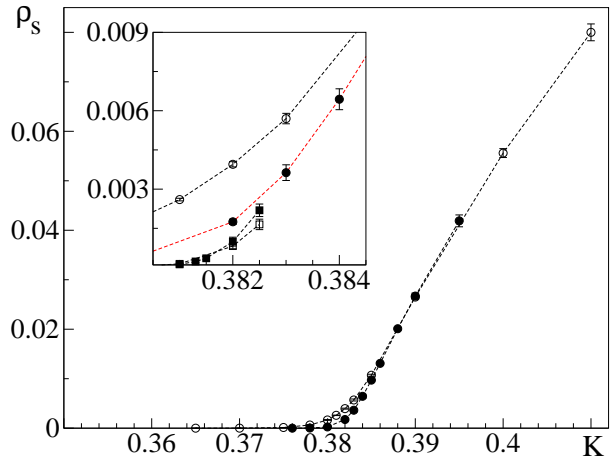}
\vspace*{-2.cm} 
\caption{ Superfluid stiffness of the broken-bond model as a function
of $K$ at different system sizes; 
$40 \times 40 \times 40$ - open  circles,  
$80 \times 80 \times 80$ - filled  circles,  
$160 \times 160 \times 160$ - open squares,  
$160 \times 160 \times 500$ -  filled  squares.  
}
\label{Fig2}
\end{figure}
\begin{figure}[tbp]
\includegraphics[width=6.5cm]{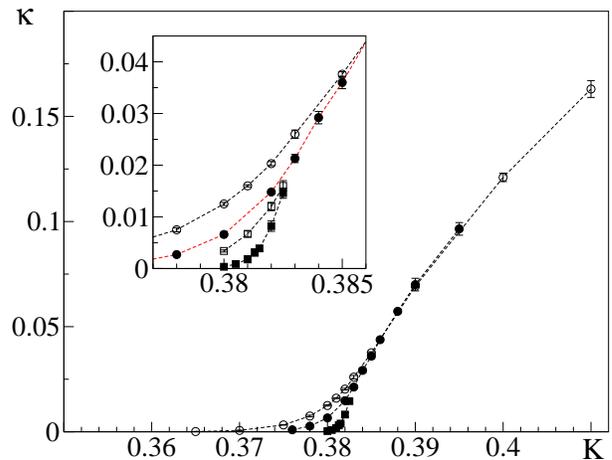}
\vspace*{-2.cm} 
\caption{ Compressibility of the broken-bond model as a function
of $K$ at different system sizes; 
$40 \times 40 \times 40$ - open  circles,  
$80 \times 80 \times 80$ - filled  circles,  
$160 \times 160 \times 160$ - open  squares,  
$160 \times 160 \times 500$ - filled  squares.  
}
\label{Fig3}
\end{figure}

We start with the off-diagonal disorder case, and consider a
system with a quarter of all bonds being broken. Typically,
we include about $10^3$ disorder realizations into the statistics
for system sizes $L\le 40$, and $ 4\times 10^4 /L$ for larger $L$.
The critical point, $K_c$, and the dynamical exponent, $z$, may be obtained
from the study of the Green function, $G({\bf r}, \tau)$, naturally 
evaluated within the WA approach \cite{Worm}. At the
critical point one should see a power-law decay: $G( r, 0) \to
r^{-(z+\eta )}$ as $r \to \infty$, and  $G( 0, \tau) \to
\tau^{-(1+\eta /z )}$ as $\tau \to \infty$. This way we find (see
Fig.~\ref{Fig1}, as well as Figs.~\ref{Fig2}, and \ref{Fig3})
\begin{equation}
K_c=0.3813(2) \; ,
\label{Kc}
\end{equation}
$z+\eta = 1.37(8)$, $1+\eta /z = 0.82(6)$, i.e.
\begin{eqnarray}
z &=& 1.65(20)  \label{z} \; ,\\
\eta &=&-0.3(1) \label{e} \; .
\end{eqnarray}
It is clear in Fig.~\ref{Fig1} that the asymptotic
behavior of the correlation function sets in only
at sufficiently large space-time distances $> 10$ lattice periods.
Moreover, the short-range behavior of $G$  mimics the
critical point of the SF-MI transition in the regular system,
where $z=1$. This peculiar behavior implies that the curves for
the superfluid stiffness, $\rho_s$,  and compressibility, $\kappa $,
will acquire their universal forms only in a very narrow region
around the critical point. Away from this region, the form of $\rho_s(K-K_c)$
and $\kappa (K-K_c)$ curves should be essentially different,
as suggested by the extended transient evolution of $z$ from $\approx 1$
to its true critical value. In Figs.~\ref{Fig2} and \ref{Fig3}
we indeed observe such a behavior. The anomalously narrow
critical region makes it virtually impossible---even with our
large cluster sizes---to reliably determine the correlation radius critical
exponent $\nu$. Along with the dynamical exponent $z$ it is
supposed to determine the critical behavior of the compressibility,
$\kappa \propto (K-K_c)^{\nu (2-z)}$ and superfluid stiffness,
$\rho_s \propto (K-K_c)^{\nu z}$ \cite{Fisher89}. The data in
Figs.~\ref{Fig2} and \ref{Fig3} at best guarantee only the inequalities
$\nu (2-z) < 1$ and $\nu z > 1$, but do not allow us to test
the Harris criterion \cite{Harris} $\nu > 2/d=1$.

The finite-size scaling of the data for compressibility
demonstrates no sign of saturation below $K_c$ and thus strongly
suggests that in the insulating state the compressibility
vanishes. Though the insulating state is incompressible, it is
easy to prove that it is {\it gapless} and thus is qualitatively
different from the conventional Mott insulator and Bose Glass
states. Indeed, in an infinite system it is always possible to
find an arbitrarily large cluster that is nearly uniform (in the
sense that fluctuations of $K$ away from its cluster average value
are arbitrarily small/rare). Taking into account that $K_c$ in the
disordered system is larger than the ideal-system critical value
$K_c^{(0)}$, we conclude that such clusters are nothing else but
finite-size {\it superfluid} lakes.  Hence, the gap associated
with adding one more particle to the cluster scales as $1/l^d$,
where $l$ is the cluster size. The absence of an upper bound on
$l$ immediately implies the absence of the global gap in the
system spectrum and finite optical conductivity.

\begin{figure}[tbp]
\includegraphics[width=6.5cm]{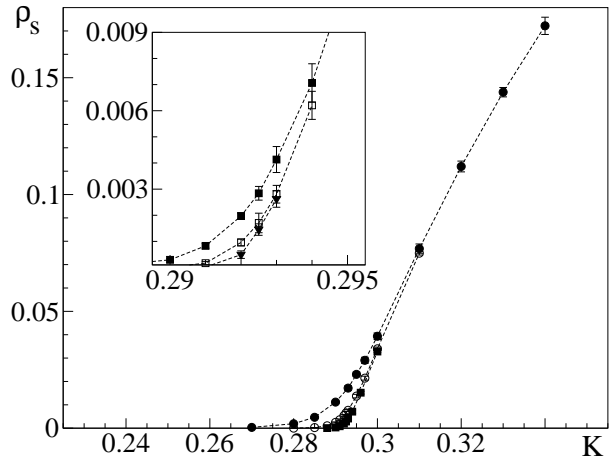}
\vspace*{-2.cm} \caption{ 
Superfluid stiffness for the diagonal disorder case 
as a function of $K$ at different system sizes; 
$10 \times 10 \times 20$ - filled circles,  
$20 \times 20 \times 49$ - open circles,  
$40 \times 40 \times 121$ - filled squares,  
$80 \times 80 \times 298$ - open squares,
$160 \times 160 \times 733$ - triangle down
}
\label{Fig4}
\end{figure}
\begin{figure}[tbp]
\includegraphics[width=6.5cm]{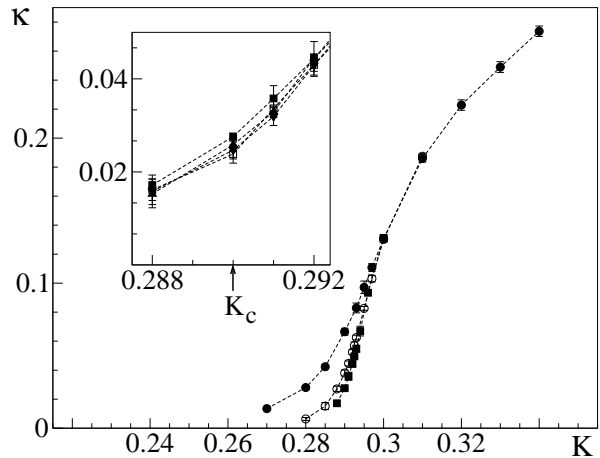}
\vspace*{-2.0cm} \caption{ 
Compressibility for the diagonal disorder case 
as a function of $K$ at different system sizes; 
$10 \times 10 \times 20$ - filled circles,  
$20 \times 20 \times 49$ - open circles,  
$40 \times 40 \times 121$ - filled squares,  
$80 \times 80 \times 298$ - open squares,
$160 \times 160 \times 733$ - triangle down,
$160 \times 160 \times 160$ - triangle up. The data for $L\ge 80$
collapse on each other within the error bars.}
\label{Fig5}
\end{figure}

The diagonal-disorder case is different. Previously reported data
\cite{Kisker} for small clusters $L\times L \le 12 \times 12$ and
disorder strength $\Delta =0.2$ were interpreted as a direct SF-MI
transition with $z=1$. We extended the study of the $\Delta =0.2$
case to system sizes $L\times L \le 160 \times 160$ and did not
find any deviations from the direct transition picture with
vanishingly small compressibility below $K_c (\Delta =0.2) =
0.325(1)$. However, the value of the MI gap in the ideal system is
almost three times smaller than $\Delta $ at $K=K_c$, see
Fig.~\ref{Fig0}. According to the argument/theorem of
Refs.~\cite{Fisher89,FM96}, the state with $\Delta > E_{\rm gap}$
is a compressible (gapless) insulator, or BG, because in
the infinite system one can always find arbitrary large regions
with the chemical potential being nearly homogeneously shifted
downwards or upwards by $\Delta$ (and thus they are doped with
particles or holes). Since the distance between such regions is
exponentially large for $\Delta \to 0$, their effect is simply
undetectably small for $\Delta =0.2$.

Even if the state right below  $K_c$ is a compressible insulator,
the question remains whether Griffiths-McCoy singularities are
inseparable from  critical fluctuations and ultimately result in
the crossover to the generic SF-BG transition, or they merely
provide a regular background contribution to $\kappa $ on which a
singular contribution $\kappa_{\rm sing}$ is superimposed. The
latter scenario implies a cusp on the compressibility curve, and
criticality different from SF-BG. To answer this question we
performed simulations for disorder strength $\Delta =0.4$. As
before, the ideal MI gap at the transition point $K_c=0.2910(5)$
is about two times smaller than $\Delta$, and $\kappa $ has to be
finite at $K_c$.

In Figs.~\ref{Fig4} and \ref{Fig5} we show the data for the compressibility
and superfluid stiffness which away from the critical point
mimic the ideal-system behavior (with the correlation length exponent
$\nu \approx 0.7$ and $z\approx 1$), but close to $K_c$ show a spectacular 
crossover to another universality class. Strong finite-size corrections to $\kappa $
for system sizes $L\le 20$ saturate for $L > 20$, and the thermodynamic curve
clearly demonstrates {\it finite}, and {\it non-singular} dependence
$\kappa (K-K_c)$. At the same time, we observe a crossover
in the $\rho_s (k-K_c)$ dependence, and see that $\rho_s$ approaches zero
with zero derivative, i.e. $\nu z >1 $.
From the decay of the Green function at the critical point we obtain 
\begin{eqnarray}
z   &=& 2.0(2)            \label{zd} \; , \\
 \eta &=&0.11(2)
\label{ed} \; .
\end{eqnarray}
Unfortunately, the large-scale crossover did not allow us
to determine the critical exponent $\nu$ from this set of data.
Recent data 
for half-integer $n$ are best fit with $\nu =1.15$, but they
also suffer from large finite-size corrections \cite{AS}. 
Apparently, the best strategy in the future is to search for a
classical model with the smallest crossover scale.

\begin{figure}[tbp]
\includegraphics[width=6.5cm]{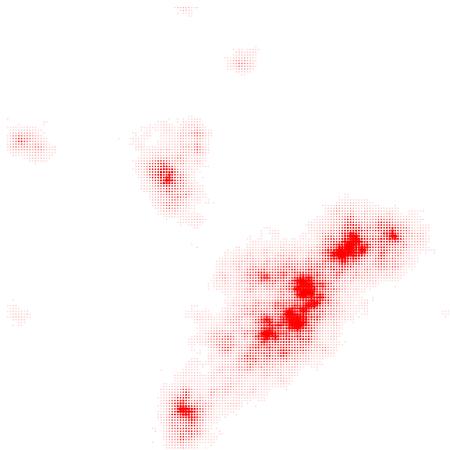}
\vspace*{-2.0cm} \caption{ A map of doped places slightly below
$K_c$ showing the picture of rare, well isolated  regions. Point
sizes are proportional to $\phi_N ({\bf r})$. } \label{Fig6}
\end{figure}

Within the WA approach one may directly visualize places where particles
are added/removed at low temperature in a sample with a given disorder realization.
The standard procedure of subtracting density maps obtained in
canonical simulations $n(i,N)-n(i,N-1)$ is time consuming, and for large systems
requires extremely high-precision data for $n(i,N)$. The new technique
is based on the statistics of the Green function calculated at the chemical
potential between the steps on the $N(\mu)$ curve, i.e. at
$\mu =E_G(N)-E_G(N-1)$, where $E_G(N)$ is the ground state energy of the
$N$-particle system. In the $T \to 0$ limit we write
$ G(N; {\bf r},{\bf r}', \tau =\beta/2) =  F(N;{\bf r},{\bf r}')$
and notice that only ground states with $N-1$ and $N$ particles contribute
to the answer. Thus
$F(N;{\bf r},{\bf r}') \approx \phi_N ({\bf r}) \phi_N ({\bf r}')$
where $\phi_N ({\bf r})$ is given by the positive-definite groundstate-groundstate
matrix element
\begin{equation}
\phi_N ({\bf r}) =  \langle \Psi_G(N)| b_{\bf r}^{\dag} | \Psi_G(N-1) \rangle \;,
\label{phi}
\end{equation}
and may be viewed as the added-particle ``wavefunction''. Its localization
length diverges at the SF-I transition.
The two important parameters which characterize the structure of
the normalized wavefunction are the localization radius
\begin{equation}
{\cal R}^2 =  \langle \; 
\left[ {\bf r} -\langle  {\bf r} \rangle \right]^2 
\, \rangle
= \sum_{\bf r} \; \left[ {\bf r}-\langle  {\bf r}
\rangle  \right]^2 \phi_N^2({\bf r}) \;,
\end{equation}
and the  state ``area'', or the number of sites over which $\phi$ delocalizes,
\begin{equation}
 {\cal A} = { \sum_{\bf r} \phi_N^2({\bf r})
 \over \sum_{\bf r}\phi_N^4({\bf r}) } \;.
\end{equation}
The dependence of ${\cal A}$ on ${\cal R}$ gives the fractal
dimension of the state. It is also important to
monitor correlations in the overlaps between different states,
$\sum_{\bf r} \phi_N({\bf r}) \phi_{N'}({\bf r})$. For example,
the percolation type scenario of the SF-I transition assumes
large fractal superfluid ``lakes'' in the
insulating phase; if so, then with the probability of order unity
there must exist an almost complete overlap between a pair of
states $(\phi_{N_a},\phi_{N_b})$ with $ N_a, N_b$ from a narrow
interval of width $\delta N \ll N $ around $N$.

In Fig.~\ref{Fig6} we show a typical map of
$\sum_{{\bf r}'} F(N=1;{\bf r}, {\bf r}')$
for the insulating state at $K=0.288$ and system size
$L \times L \times L_{\tau} =160 \times 160 \times 1000 $. One
can clearly see isolated regions doped with particles.

In summary, we have performed large-scale simulations of the
superfluid--insulator transition in the $(2+1)$-dimensional classical
analog of the commensurate disordered 2D bosonic system.
For diagonal disorder, our results suggest that commensurability
is not relevant in the long-range limit, and the universality class
of the transition (superfluid--Bose-glass) is the same for all
filling factors. In particular, we unambiguously resolved the
finite compressibility at the critical point originating from
rare statistical fluctuations and demonstrated that they are responsible
for the crossover in critical behavior. We believe that our results
rule out the earlier-reported direct superfluid--Mott-insulator
transition in this model.

In the off-diagonal disorder case, the compressibility vanishes
at the critical point. The incompressible insulating phase,
however, is gapless, and its universality class is characterized by
the dynamical critical exponent $z=1.65 \pm 0.2$.

A general observation is that even for large diagonal and
off-diagonal disorder, the universal asymptotic long-range behavior
sets in only at large space-time distances ($\sim 20$
lattice periods). This circumstance explains previous observations
of the direct superfluid--Mott-insulator transition in small-size
clusters and implies that the the superfluid stiffness and
compressibility should obey generic scaling laws only in a very
close vicinity of the phase transition point.

The authors are grateful to S.~Sachdev for a fruitful discussion.
This work was supported by the National Science Foundation under
Grant DMR-0071767. BVS acknowledges a support from Russian
Foundation for Basic Research under Grant 01-02-16508, from the
Netherlands Organization for Scientific Research (NWO), and from
the European Community under Grant INTAS-2001-2344.


\end{document}